# Frequency of Occurrence and Information Entropy of American Sign Language

Andrew Chong, Lalitha Sankar, H. Vincent Poor
Princeton University, Princeton, NJ 08554.

## Abstract

American Sign Language (ASL) uses a series of hand based gestures ("signs") as a replacement for words to allow the deaf to communicate. Previous work has shown that although it takes longer to make signs than to say the equivalent words, on average sentences can be completed in about the same time. This leaves unresolved, however, precisely why that should be the case. This paper reports a determination of the empirical entropy and redundancy in the set of *handshapes* (conformation of the hand) of ASL. In this context, the entropy refers to the average information content in a unit of data. It is found that the handshapes, as fundamental units of ASL, are less redundant than phonemes, the equivalent fundamental units of spoken English, and that their entropy is much closer to the maximum possible information content (achieved by equally likely handshapes). This explains why the slower signs can produce sentences in the same time as speaking – the low redundancy compensates for the slow rate of sign production. In addition to this precise quantification, this work is also novel in its approach towards quantifying an aspect of the ASL alphabet. Unlike spoken and written languages, frequency analysis of ASL is difficult due to the fact that every sign is composed of phonemes that are created through a combination of manual (hand) and a relatively large and imprecise set of bodily features. Focusing on handshapes as the ubiquitous and universal feature of all sign languages permits a precise quantitative analysis. As interest in visual electronic communication explodes within the deaf community, this work also paves the way for more precise automated sign recognition and synthesis.

# Frequency of Occurrence and Information Entropy of American Sign Language

## Introduction

American Sign Language (ASL) is a hand-based gestural language used primarily by about 2 million deaf and hard-of-hearing Americans [4]. Series of hand-based gestures ("signs") are used in ASL as replacement for words. Signs are composed of phonemes, which in turn are created through a combination of *features* such as the conformation of a hand, commonly known as a *handshape*, the movement of a hand or hands, and the location of the hands with respect to the body, and other non-manual physical actions such as smiling, shrugging, and nodding.

The handshape feature is of particular interest to researchers because of their ubiquitous nature in signed language: each sign includes at least one handshape. There is a discrete number of handshapes that are precisely defined in ASL. In contrast, other variables of signs such as palm orientation and hand movement are defined differently based on the analysis applied to ASL, i.e., one or more such variables may be defined as parts of a single feature or sometimes as individual features [1, 16]. Due to the ambiguity associated in defining these variables, in this study, we focus on handshapes as the universal feature of signed languages. For brevity, we henceforth refer to the set of handshapes as the *ASL handshape alphabet*.

In their seminal work in [6], Bellugi and Fischer showed that although English words are spoken at a faster rate than ASL signs are produced, equivalent sentences in both languages take approximately the same amount of time to produce. In this work, we offer an explanation for this effect by quantifying the predictability and redundancy of the ASL handshape alphabet. To this end, we determine the average information content in bits, i.e., the *entropy*, of the set of ASL handshapes, and compare it with the entropy of the set of phonemes in the English language. The motivation for choosing the two sets comes from the fact that in both languages, the elements of these sets are the fundamental units of the respective languages [ref. for phonemes in English and ASL].

We compute the entropy of the ASL handshape alphabet by determining the empirical frequencies of the handshapes. While specific ASL studies have targeted the frequencies of certain classes of signs, to the best of our knowledge, empirical frequencies of handshapes have not been computed prior to this work. We use signed video logs and recorded conversations to obtain the empirical frequencies. Our analysis reveals that the empirical entropy for handshapes, in comparison to the entropy for phonemes in English [12] is much closer to its maximum possible entropy (achieved by an alphabet with equally likely letters), and that ASL handshapes are relatively less redundant than English phonemes. Our observations lead us to conclude that, as a universal feature of all signs, handshapes have lower redundancy to compensate for the slow rate of sign production.

## Information Theory Concepts

In his paper "A Mathematical Theory of Communication", Claude Shannon defined entropy as a measure of the average information content of an information source [2]. A distinction between different sources of information, such as images, videos, audio, or text, can be made using the alphabet from which the source takes values. Thus, for example, the alphabet of images is a set of RGB (red, green, and blue) pixel values, while that for text is the alphabet of the language used. Noting that a deterministic source, i.e., an information source that presents the same letter of the alphabet at all times provides no information over time, Shannon modeled an information source as a random variable. For our purposes, a random variable $X$ is quantified by a finite alphabet $\mathcal{X}$ from which it takes values $x$ with probability (relative frequency of occurrence) $p(x)$ such that the probabilities of all elements of the alphabet sum to one. Using this model, Shannon defined the information content (in units of bits) of each letter of the alphabet as $H(x) = \log_2(1/p(x))$ and thus, the entropy $H(X)$ as the weighted sum of $H(x)$ for all $x$ in $\mathcal{X}$, where the weight for each letter $x$ is its probability $p(x)$. Entropy defined in this way also quantifies the average uncertainty of an information source. Thus, among all sources taking exactly $N$ values, a source whose alphabet consists of $N$ equally likely letters has the most uncertainty about it and a maximum entropy of $\log_2(N)$. On the other hand, a source with a distinctly biased alphabet is more predictable and has a lower average entropy relative to the maximum.

In his seminal paper on the entropy and predictability of the English language [3], Shannon determined the empirical entropy of the English language for a 27-letter alphabet (with whitespace as the $27^{th}$ character). The empirical entropy is computed in the same way as the entropy except that the empirical frequencies of the letters are used in the place of the probabilities. Shannon defined the redundancy $R$ as the difference from 1 of the ratio of the empirical entropy to the maximum entropy where the maximum entropy is obtained by assuming that all the letters of the alphabet are equally likely. This ratio quantifies the "maximum degree of compression" possible when the language is encoded to the same alphabet while, its difference from 1, i.e., the redundancy, quantifies the predictability of the language. For printed English, the redundancy has been shown to be in the neighborhood of 50%, indicating that nearly half of a text can be predicted when the remaining half is known. A very low value for redundancy indicates that a large number of letter combinations are allowed in the language whereas a very high redundancy indicates that the language is heavily constrained.

## Spoken versus Signed Language

ASL, as a gestural language, differs significantly from English and, in fact, all spoken languages. ASL is transmitted by hand gestures and is received visually, while conversational English is transmitted by vocal cord vibrations and is received aurally. ASL is almost always used in person-to-person communication, and rarely written, as there is no single well-established or accepted system for writing in ASL. ASL is also word-based, in that there are no distinct "letters" – when signed, gestures are used to represent words or phrases, and spelling by letters, known as *fingerspelling*, is used only in special situations.

In linguistic terminology, a phoneme is the smallest structural unit capable of conveying a distinction in meaning [8]. For example, the /m/ in mat and /b/ in bat are phonemes. Phonemes in American Sign Language are the basic elements of gesture and location that are combined to create individual signs in ASL, and are generally considered to be linguistically identical to phonemes in written languages [6]. ASL phonemes are in general distinguished based on multiple features such as handshape, location of the sign on the body, palm orientation, movement of the hands, and non-manual features such as facial expression or head movement. Features such as movement and palm orientation may be considered indistinguishable depending on the linguistic analysis applied to ASL. Furthermore, these features are often difficult to study – for example, there are so many possible locations that it is not easy to establish frequencies of use for each location. Palm orientation has too few variations and is not always used in every sign. Movement has too many variations and does not appear in every sign. Finally, non-manual features are in general too hard to categorize and are also not used in every sign.

Handshapes are the only type of feature present in all signs. There are approximately 150 different handshapes, with about 41 phonemically distinct handshapes [11, 17]. Futhermore, handshapes are the only feature recognized by all major ASL linguistic analyses. For these reasons, we focus here on the set of handshapes as a collection of fundamental units of the language.

**<u>Materials and Methods</u>**

We compiled a list of handshapes using [1, 4, 10, 11, 17]. It has been estimated that there are about 45 different handshapes [16]. However, not all of these handshapes are phonemically distinct [1, 17], as some handshapes are interchangeable and thus redundant. Of the 45 handshapes, 23 have a one-to-one correspondence with 23 distinct letters of the English alphabet. The handshapes for the remaining three letters, I, K, and D, are the same handshapes as those for J, P, and Z, respectively. In signed conversations, these letters are distinguished based on palm orientation and movement that we ignore for the purpose of this analysis. Furthermore, there are 10 distinct handshapes used to represent the digits 0 through 9. However, the handshapes representing the letters D, V, W, and F are the same as those for the numbers 1, 2, 6, and 9. In signed conversation, the distinction is made contextually and through the use of other sign features that we ignore for the purpose of this analysis. In addition to these 29 handshapes that have an alphanumeric correspondence, there are 16 handshapes used for signing. The complete set of handshapes is given by the following list. We note that in the list below the names for handshapes are not universal; in fact, some of the handshapes may have several names associated with them depending on the signs in which they are commonly used.

- A / 10: fist with thumb touching index finger
- Open A: fist with thumb extended
- B: flat hand with fingers together
- Open B: flat hand with fingers together, thumb apart
- Bent B: flat hand with thumb apart, bent at first joint
- C: curved hand, with thumb apart
- D: lotus hand with pointing index finger
- E: curled fingers with thumb along nails
- F: thumb touching index finger, other fingers straight
- Open F: thumb almost touching index finger, other fingers slightly curled
- G: thumb and index finger extended, other fingers curled to palm
- H: index and middle finger extended, other fingers curled to palm
- I: fist with pointing pinky finger
- K/P: victory sign with thumb in between index and middle fingers
- L: fist with thumb and index finger apart
- Bent L: thumb and index finger extended at right angle, other fingers curled
- M: thumb tucked in fist between ring and pinky fingers
- N: thumb tucked in fist between ring and middle fingers
- Open N: thumb, index, and middle fingers extended
- O: thumb touching fingertips
- Baby O: thumb and index touching, other fingers curled in fist
- Flattened O: thumb touching other fingers, all mostly flattened
- R: fist with crossed index and middle fingers
- S: fist with thumb folded on index and middle finger

- T: fist with thumb between index and middle finger
- U: fist with pointed index and middle fingers together
- V: fist with pointed index and middle fingers apart
- Bent V: thumb touching ring and pinky fingers, index and middle fingers bent V
- W: fist thumb touching pinky
- X: fist with hooked index finger
- Y: fist with pinky and thumb extended
- L-I / ILY: thumb, index, and pinky fingers extended, other fingers curled to fist
- 1-I / Y: Pinky and index finger extended, thumb touching middle and ring fingers
- 1: index finger extended, other fingers in fist
- 3: fist with thumb, index, and middle finger extended
- Bent 3: thumb, index, and middle fingers extended, other fingers in fist
- 4: flat hand with fingers spread
- 5: spread hand
- Bent 5: all fingers bent and apart
- 8: index and middle finger touching, other fingers extended
- Open 8: flat hand with middle finger bent

We collected data using video logs (vlogs) from on-line sources such as YouTube.com deafvideo.tv, and deafread.com. Video logs have become very popular in the Deaf community, due to the rise of fast Flash-based video sites like YouTube.com and due to the fact that vlogs allow signers to communicate visually. Vlogs were selected for at least 30 seconds of continuous speech. In addition, we also used signed conversations that we recorded (videotaped) at the New Jersey School for the Deaf, Katzenbach Campus. To allow for diversity in conversations, and thus, a diversity in the use of handshapes, conversations of students of high school age or older and teachers at the Katzenbach Campus were videotaped over a period of two months during classroom discussions, class breaks, and lunch breaks.

The relative frequency of each handshape is computed by counting the number of times a handshape appears in the entire conversation or vlog. Thus, a sign can include many repetitions of a handshape and each such occurrence is counted. The relative frequency $f(x)$ of a handshape $x$ is obtained by dividing the total number of occurrences of tgat handshape by the total number of handshapes. The information in bits of each handshape is given by $\log_2(1/f(x))$, where $f(x)$ is the relative frequency (empirical probability) of $x$. These individual handshape frequencies are calculated for the two sets of data. A plot of the handshape frequencies for the vlogs and the conversations is shown in Fig. 1. For both sources, the information content of each handshape is shown in Fig. 2.

The average entropy was calculated using (formula reference) and was found to be approximately 5.14 bits for vlogs and 5.01 bits for conversational signing. For comparison, the maximum possible entropy for handshapes is $\log_2(45) = 5.49$ bits. In contrast, the results in [12] indicate that the empirical entropy of phonemes is about 0.6-1.5 bits. Thus, we see that difference between the empirical entropy and the maximum entropy of the ASL handshape alphabet is within 0.5 bits. In contrast, the empirical entropy for English phonemes is about 3 to 3.5 bits lower than the maximum possible entropy.

**Discussion**

Our findings suggest that a slow rate of sign production in ASL may be compensated for, at least in part, by a low redundancy of handshapes. This offers an explanation for the effect that Bellugi and Fischer discovered in their comparison of English and ASL, i.e., that sentences can be produced at roughly the same rate as in spoken English, despite the relatively slow rate of signing [6].

Entropy might be higher for handshapes than English phonemes because the visual channel is less noisy than the auditory channel. Environmental sounds might cause interference when visual effects might not. This interference might be common enough that English speech relies on redundancy of phonemes in order to correct errors introduced by this interference. By this logic, entropy might be higher in ASL handshapes because conversely the visual channel as used by sign language is inherently less noisy than the auditory channel as used by spoken language, and so error correction is less necessary.

Error correction in ASL might be accomplished not with handshape redundancy but by holding handshapes for longer periods of time. Difficulties in visual recognition of handshapes could be solved by holding or slowing the transition between those handshapes for longer amounts of time, while difficulties in auditory recognition of spoken phonemes cannot always be easily solved by speaking phonemes for longer amounts of time.

Other ASL features may provide more information for error correction or reducing redundancy even further. It is important to remember that handshapes are not actually phonemes, although we have compared handshapes to English phonemes. The phoneme set might be expanded greatly through the use of other sign features (location, palm orientation, movement, and non-manual features). These other sign features might also be redundant in their own frequencies or in their combination with handshapes. We suspect specifically that the combinations of handshapes and motions between the dominant and non-dominant hands are redundant to some degree, and that this might provide error correction where simple handshape frequency does not. Further analysis in terms of entropies for the other features and mutual information of concurrent handshapes may reveal such effects.

In addition to recorded signed conversations, we used video logs for our analysis. Video logs have enabled the Deaf community to participate in the new age of electronic communications. In contrast to the low rate transmission required of spoken communication via devices such as cellular phones, signed communications require higher bandwidth medium to facilitate video transmissions. In fact, the *MobileASL* group at the University of Washington is developing video compression schemes to enable real-time reliable signed communications for handheld devices over the wireless medium [21]. Our frequency and redundancy analysis in combination with additional analysis of related ASL features can result in the design of better encoders and decoders that achieve compression rates required for wireless transmissions.

**References**

1. W.C. Stokoe, Dictionary of American Sign Language on Linguistic Principles. Linstok Press, Inc., 1976.
2. C. E. Shannon, "*A Mathematical Theory of Communications*", Bell System Technical Journal*, pages 379-423, 1948.
3. C.E. Shannon, "*Prediction and Entropy of Printed English*", Bell System Technical Journal, pages 50-64, January 1951.
4. National Institute on Deafness and Other Communication Disorders, "*American Sign Language*", http://www.nidcd.nih.gov/health/hearing/asl.asp
5. S.D. Fischer, L.A. Delhorne, & C.M. Reed, "*Effects of rate of presentation on the visual reception of American Sign Language*", Journal of Speech, Language, and Hearing Research, http://www.ncbi.nlm.nih.gov/pubmed/10391623, 1998.
6. U. Bellugi & S. Fischer, "*A comparison of sign language and spoken language*", Cognition, 1, pages 173-200, 1972.
7. "*American Sign Language*", http://en.wikipedia.org/wiki/American_Sign_Language, 2008.
8. SIL International, "*What is a phoneme?*", http://www.sil.org/linguistics/GlossaryOfLinguisticTerms/WhatIsAPhoneme.htm, 2004.
9. P. Eccarius & D. Brentari, "*Symmetry and dominance: A cross-linguistic study of signs and classifier constructions*", Lingua, Volume 117, Issue 7, July 2007, Pages 1169-1201.
10. M.L.A. Sternberg, American Sign Language Dictionary, Collins, 1998.
11. Random House Webster's Unabridged American Sign Language Dictionary, Random House, ed. E. Costello, 2001.
12. V. Van de Laar, B. Kleijn, & E. Deprettere, "*Perceptual entropy rate estimates for the phonemes of American English*", IEEE International Conference on Acoustics, Speech, and Signal Processing, Volume 3, page 1719, 1997.
13. J.H.D. Allen, Jr., L.S. Hultzen, & M.S. Miron, "*Tables of transitional frequency of English phonemes*", Linguistic Society of America, Vol. 41, No. 3, pages 525-529, 1965.
14. J.B. Plotkin, & M.A. Nowak, "*Language Evolution and Information Theory*", Journal of Theoretical Biology, Volume 205, Number 1, pages 147-159, July 2000.
15. J. Ann, Frequency of Occurrence and Ease of Articulation of Sign Language Handshapes: The Taiwanese Example. Gallaudet University Press, 2006.
16. R. Battison, "*Analyzing Signs*", Lexical Borrowing in American Sign Language, Linstock Press, 1978.
17. R. A. Tennant, M.G. Brown. The American Sign Language Handshape Dictionary. Gallaudet University Press, 1998.
18. T. Burger, A. Benoit, A. Caplier, "Intercepting Static Hand Gestures in Dynamic Context". International Conference on Image Processing, 2006.
19. J. Morford, J. MacFarlane. "Frequency Characteristics of American Sign Language". Sign Language Studies, Vol. 3, No. 2, pages 213-225, 2003.
20. U. Bellugi & S. Fischer, "*A comparison of sign language and spoken language*", Cognition, 1, pages 173-200, 1972.
21. A. Cavender, R. LAdner, E. Riskin, "MobileASL: intelligibility of Sign Language video over mobile phones," Disability and Rehabilitation: Assistive Technology, London: Taylor and Francis, pp 1-13, June 2007.

**Figures**

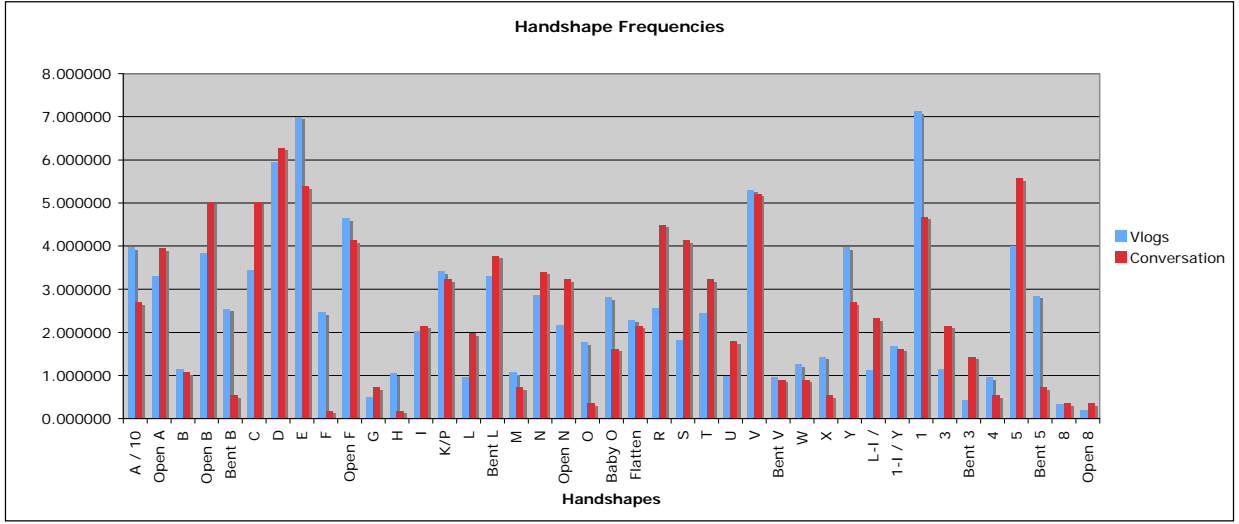

**Figure 1: Handshape Frequency**

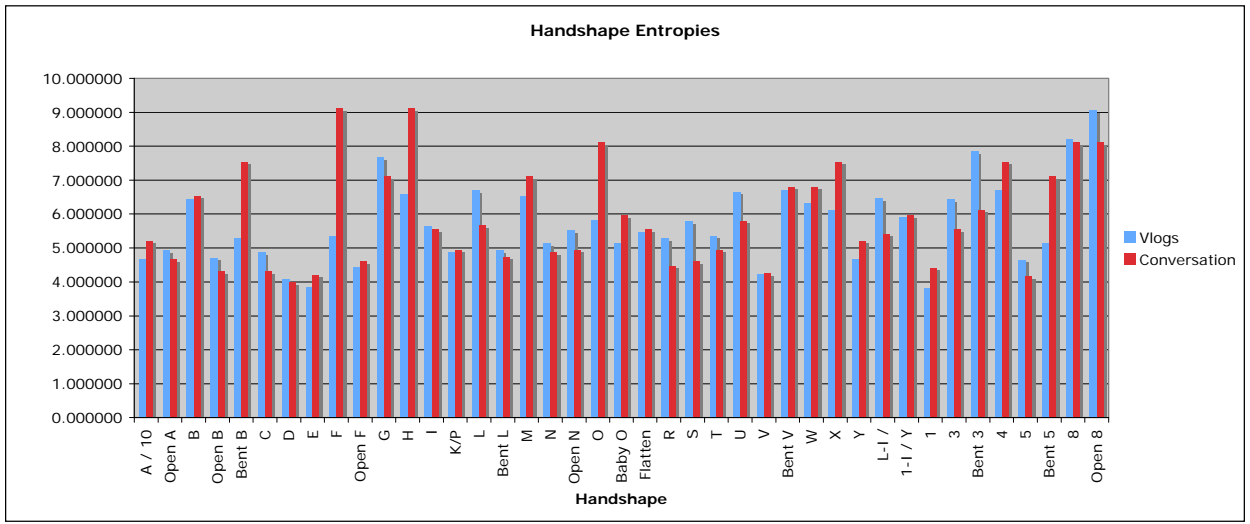

**Figure 2: Handshape Entropy**